\documentclass[prb,reprint,superscriptaddress,a4paper,aps,showpacs]{revtex4-1}
\usepackage{graphicx}\graphicspath{{./},{/home/user/fig/publi/muon/}}
\usepackage{psfrag}
\usepackage{amsmath}

\begin{document}

\title{Nd$_2$Sn$_2$O$_7$: an all-in--all-out pyrochlore magnet with no divergence-free field and anomalously slow paramagnetic spin dynamics}

\author{A. Bertin}
\author{P. Dalmas~de~R\'eotier}
\author{B. F{\aa}k}
\author{C. Marin}
\author{A. Yaouanc}
\affiliation{Univ.\ Grenoble Alpes, INAC-SPSMS, F-38000 Grenoble, France}
\affiliation{CEA, INAC-SPSMS, F-38000 Grenoble, France}
\author{A.~Forget}
\affiliation{Institut Rayonnement Mati\`ere de Saclay, SPEC, CEA,
F-91191 Gif-sur-Yvette, France}
\author{D. Sheptyakov}
\affiliation{Laboratory for Neutron Scattering and Imaging, Paul Scherrer Institute, CH-5232 Villigen-PSI, Switzerland}
\author{B. Frick}
\author{C. Ritter}
\affiliation{Institut Laue-Langevin, Boite Postale 156X, F-38042 Grenoble Cedex 9, France}
\author{A. Amato}
\author{C. Baines}
\affiliation{Laboratory for Muon-Spin Spectroscopy, 
Paul Scherrer Institute, CH-5232 Villigen-PSI, Switzerland}
\author{P.J.C. King}
\affiliation{ISIS Facility, Rutherford Appleton Laboratory, Chilton, Didcot, OX11 0QX, UK}

\date{\today}

\begin{abstract}

We report measurements performed on a polycrystalline sample of the pyrochlore compound Nd$_2$Sn$_2$O$_7$. It 
undergoes a second order magnetic phase transition at $T_{\rm c}  \approx 0.91 $~K to a 
noncoplanar all-in--all-out magnetic structure of the Nd$^{3+}$ magnetic moments. The thermal 
behavior of the low temperature specific heat fingerprints excitations with linear dispersion in a 
three-dimensional lattice. The temperature independent spin-lattice relaxation rate measured below $T_{\rm c}$ and
the anomalously slow paramagnetic spin dynamics detected up to $\approx 30 \, T_{\rm c}$ are suggested 
to be due to magnetic short-range correlations in unidimensional spin clusters, i.e., spin loops. The observation of a spontaneous field in muon 
spin relaxation measurements is associated with the absence of a divergence-free field for the ground state of an all-in--all-out pyrochlore magnet as predicted recently.

\end{abstract}


\maketitle

\section{Introduction}
The pyrochlore insulator compounds of generic chemical formula $R_2M_2$O$_7$, where $R$ is a rare earth 
ion and $M$ a non magnetic element, are a fertile playground for the discovery of exotic magnetic 
properties.\cite{Gardner10} The most enigmatic results were found for their spin dynamics. 
At least two signatures of them have been recognized.

First, persistent spin dynamics as inferred from a finite spin-lattice relaxation rate measured at low 
temperature by the zero-field muon spin relaxation ($\mu$SR) technique is an ubiquitous physical 
property of the ground state, no matter its magnetic nature. Low energy unidimensional excitations 
have recently been argued to be at its 
origin.\cite{Yaouanc15} They could be supported by spin loops, i.e., peculiar short-range correlations.\cite{Villain79,Hermele04}

In $\mu$SR measurements, the signature of a transition to a magnetically ordered state is the observation of muon spin precession corresponding to a spontaneous field at the muon site. While in neutron scattering experiments at low temperature, Er$_2$Ti$_2$O$_7$, Tb$_2$Sn$_2$O$_7$, and Yb$_2$Sn$_2$O$_7$ display magnetic Bragg reflections characterized by a propagation wave vector ${\bf k} = (0,0,0)$ (Refs.~\onlinecite{Champion03,Mirebeau05,Yaouanc13}) no spontaneous precession is observed in $\mu$SR measurements.\cite{Lago05,Dalmas06,Yaouanc13} This is the second unexpected result. In contrast, spontaneous fields have been reported for Gd$_2$Ti$_2$O$_7$.\cite{Yaouanc05a} However, this is a system with a complex magnetic structure.\cite{Stewart04} More interesting, Gd$_2$Sn$_2$O$_7$ displays a ${\bf k} = (0,0,0)$  structure\cite{ Wills06} and spontaneous fields.\cite{Bonville04a,Chapuis09b} A fraction of the $\mu$SR signal is missing in the ordered state, probably because of the large field distribution at the magnetic sites corresponding to this fraction. In this context the characterization of a compound with a ${\bf k} = (0,0,0)$ magnetic order together with a spontaneous field with no missing fraction would provide further insight. 

Here we report a study of the cubic pyrochlore stannate Nd$_2$Sn$_2$O$_7$ (Refs.~\onlinecite{Subramanian83,Kennedy97}) with 
specific heat, magnetization, neutron scattering, and $\mu$SR measurements. It exhibits a magnetic phase transition 
at $T_{\rm c} \approx0.91$~K to a so-called all-in--all-out magnetic structure with a  ${\bf k} = (0,0,0)$ propagation wavevector. A  $\mu$SR spontaneous field is 
observed, consistent with the lack of the divergence-free part of the Helmholtz decomposition of the magnetic-moment 
field for such a magnetic structure.\cite{Brooks14} In this frame, the long-range order is associated with the divergence-full 
component of the field. Persistent spin dynamics below $T_{\rm c}$ and an anomalously slow paramagnetic spin dynamics up to $\approx 30 \, T_{\rm c}$ are detected. They are suggested to be due to unidimensional spin loops.

We shall first establish the basic physical characteristics of the compound from bulk measurements and neutron scattering experiments. Further microscopic information on the static and dynamical properties of the system is obtained from inelastic neutron scattering and $\mu$SR experiments. A discussion of the results completes this report.
 
\section{Bulk measurements}
\begin{figure}
\begin{picture}(255,98)
\put(95,55){(a)}
\put(-5,0){
\includegraphics[height=0.145\textheight]{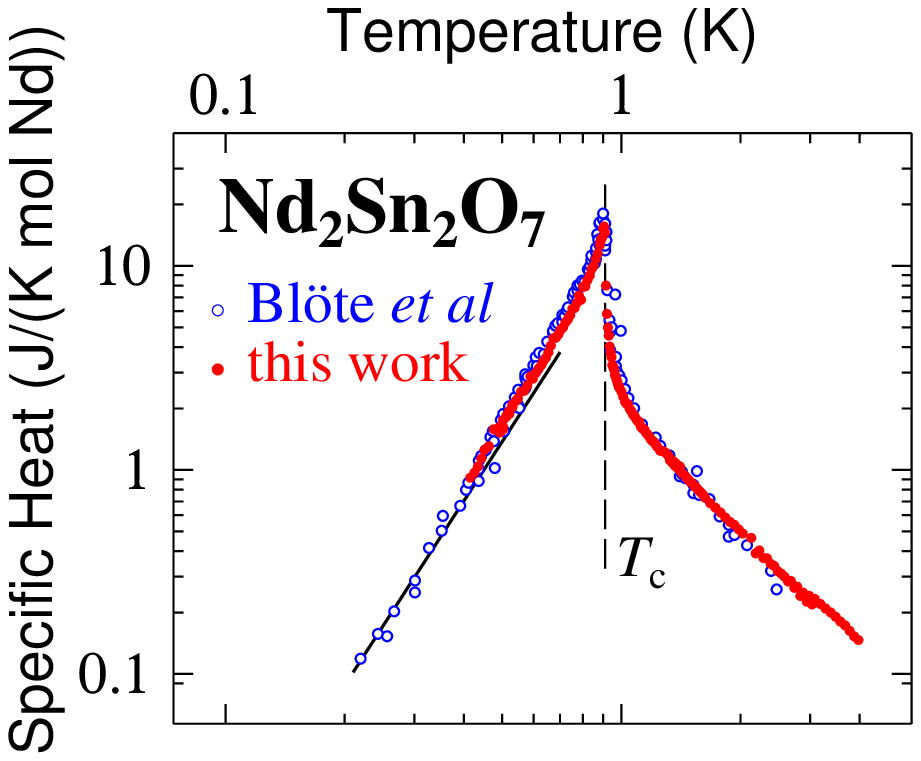}
}
\put(220,55){(b)}
\put(115,0){
\includegraphics[height=0.145\textheight]{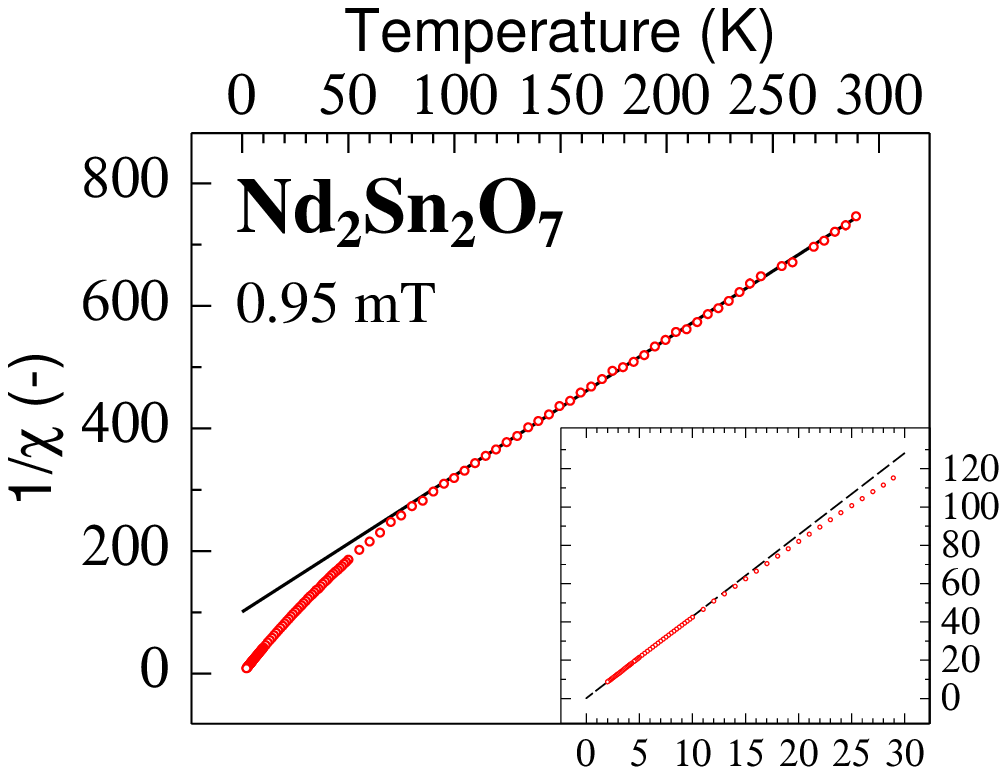}
}
\end{picture}
\caption{(Color online) Bulk measurements for a Nd$_2$Sn$_2$O$_7$ powder sample.
(a) Specific heat versus temperature. Our results are compared with those in Ref.~\onlinecite{Blote69}.
(b) Temperature dependence of the inverse magnetic susceptibility $1/\chi$ expressed in SI units and measured in a field of 0.95~mT. The insert displays the low temperature data. The sample used for this measurement was a thin pellet and the field was applied in its plane so as to minimize demagnetization field effects. 
In the two panels, the solid lines are results of fits as explained in the main text.
} 
\label{sh_T}
\end{figure}

A single phase powder sample of Nd$_2$Sn$_2$O$_7$ was prepared by heating the constituent oxides up to 1400$^\circ$C with four intermediate grindings. 
The heat capacity $C_{\rm p}$ recorded with a Physical Property Measurement System, Quantum Design Inc, is presented in Fig.~\ref{sh_T}(a). 
It matches previous results \cite{Blote69} in the overlapping temperature range. A $\lambda$-type peak consistent with a second order phase transition is observed at $T_c\approx 0.91$~K, i.e., the temperature of a peak in magnetic susceptibility data.\cite{Matsuhira02} No broad hump above $T_{\rm c}$ is present unlike in some geometrically frustrated magnets where it is interpreted as a signature of short-range correlations.\cite{Dalmas03} The law $C_{\rm p} =  {\mathcal B} T^3$ accounts for the data at low temperature [Fig.~\ref{sh_T}(a)]. This temperature dependence is expected for excitations with a linear dispersion relation in a three-dimensional system. It sets a higher bound for any gap in the excitation spectrum at $\approx 0.1$~meV. 
From the value ${\mathcal B}=11.0 \, (7)$~J\,K$^{-4}$\,mol$^{-1}$ obtained from data measured below 0.45~K, we infer an excitation velocity $v_{\rm ex}=55 \, (1)$~m\,s$^{-1}$ in line with the Er$_2$Ti$_2$O$_7$ result.\cite{Dalmas12}

In Fig.~\ref{sh_T}(b) is displayed the inverse of the static susceptibility $1/\chi$. 
In the temperature range $150 \leq T \leq 290$~K $\chi$ follows a Curie-Weiss law 
with a Curie-Weiss temperature $\theta_{\rm CW}=-46.3 \ (1.9)$~K and a paramagnetic 
moment $m_{\rm pm}=3.57 \, (4)~\mu_{\rm B}$ comparable with the value 
$m_{\rm pm} =3.62~\mu_{\rm B}$ for a free Nd$^{3+}$ ion. As shown in the insert,
assuming $\chi$ to follow a Curie-Weiss law for $5 \leq T \leq 15$~K we get 
$\theta_{\rm CW}=-0.32 \ (1)$~K and $m_{\rm pm} =2.63 \, (3)~\mu_{\rm B}$, in agreement 
with Ref.~\onlinecite{Bondah01}.
The negative $\theta_{\rm CW}$ indicates net antiferromagnetic exchange interactions. 
As the first excited crystal-field doublet is located at $\approx 26$~meV above the 
Kramers doublet of the  Nd$^{3+}$ ground-state,\cite{Bertin15a} an effective 
spin $S'$ = 1/2 model is justified for the ion description at low temperature. 

\section{Neutron diffraction results}
\begin{figure}
\begin{picture}(255,215)
\put(17,200){(a)}
\put(-4,110){
\includegraphics[scale=0.50]{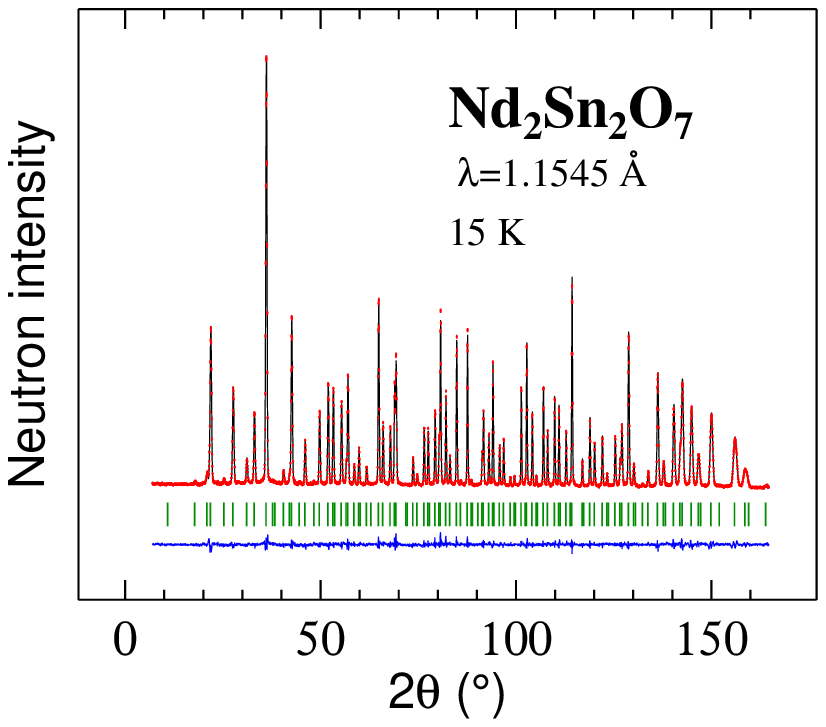}
}
\put(228,200){(c)}
\put(140,125){
\includegraphics[scale=1.00]{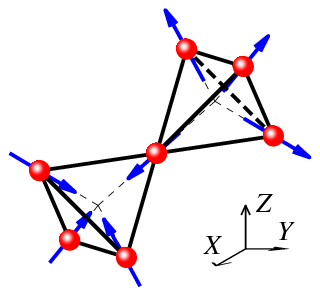}
}
\put(17,90){(b)}
\put(-4,0){
\includegraphics[scale=0.50]{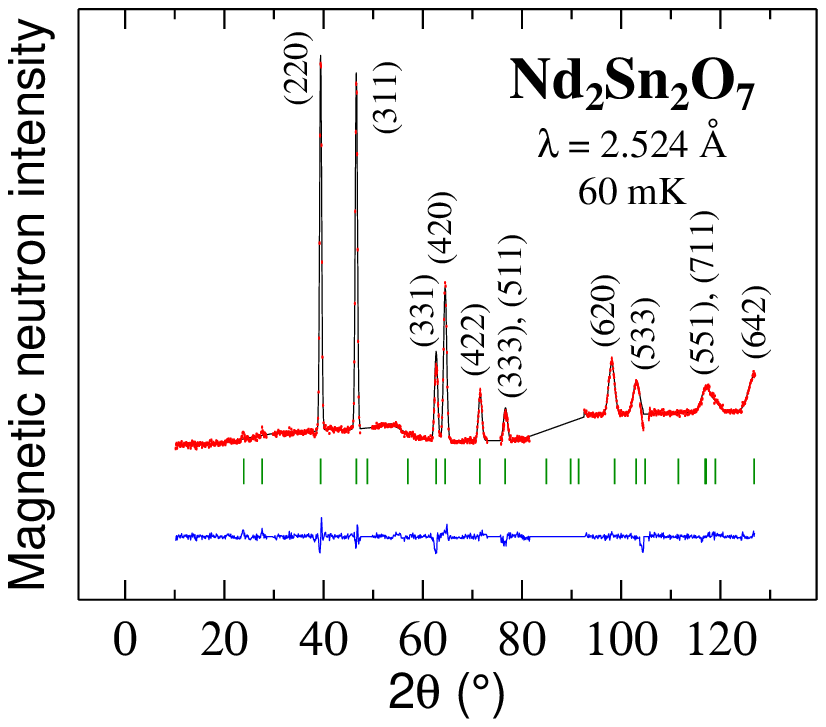}
}
\put(226,90){(d)}
\put(120,0){
\includegraphics[scale=0.50]{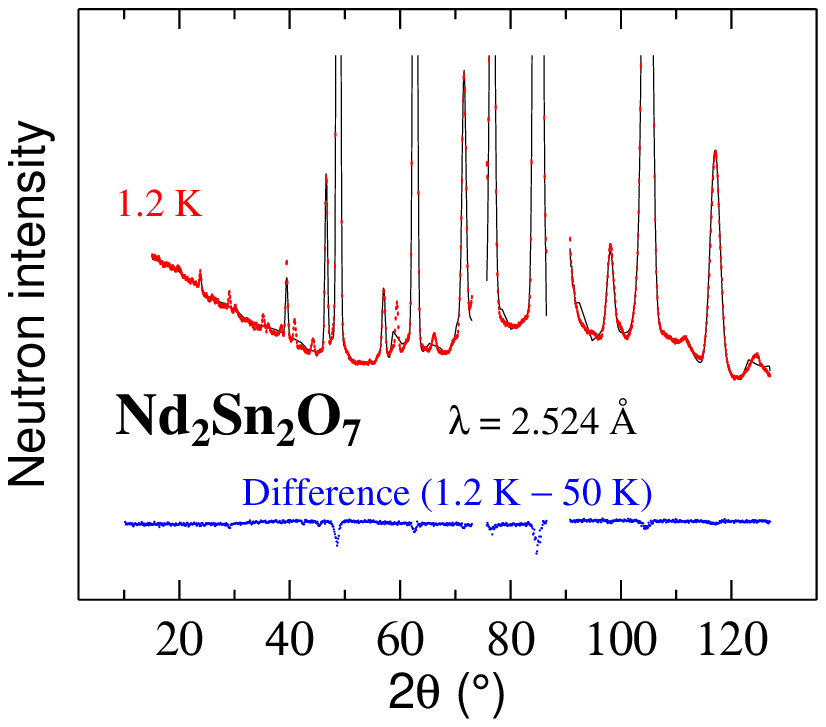}
}
\end{picture}
\caption{(Color online) Neutron diffraction results for a powder sample of Nd$_2$Sn$_2$O$_7$.
(a) Nuclear pattern at $T$ = 15~K. 
(b) Magnetic diagram given by the difference between 60~mK and 1.2~K data sets. Experimental points nearby $2 \theta$ = 74.4$^{\circ}$ and 88.5$^{\circ}$ are not shown because they are strongly influenced by neutrons scattered from the copper container.  The observed magnetic reflections are labeled with Miller indices. The solid lines in (a-b) result from Rietveld analyses. The solid lines at the bottom give the difference between fits and data. The vertical markers indicate the positions of the Bragg peaks.
(c) The corresponding magnetic structure. The spheres represent the Nd$^{3+}$ ions and the arrows their magnetic moments oriented along the local trigonal ${<}111{>}$ axes of the cubic crystal structure. Two corner-sharing tetrahedra are shown, one with the magnetic moments pointing inwards and an adjacent tetrahedron with moments pointing outwards.
(d) Diffractogram recorded at 1.2~K and difference from the data set recorded at 50~K. No diffuse scattering intensity is observed at the approach of the magnetic transition. Note that for the sake of clarity, the most intense Bragg peaks present at 1.2~K have been truncated.
}
\label{Neutron_results}
\end{figure}

Nd$_2$Sn$_2$O$_7$ crystallizes in the $Fd\bar{3}m$ space group. We studied the crystal structure of our sample using the diffractometers D2B at Institut Laue Langevin (ILL) and HRPT of the SINQ neutron source at the Paul Scherrer Institute (PSI) respectively at room temperature and 15~K. Figure~\ref{Neutron_results}(a) 
displays the HRPT data analyzed with the FullProf suite.\cite{Rodriguez93} The lattice parameter is $a_{\rm lat}=10.5586\,(6)$~\AA\ and the parameter for the oxygen position is $x=0.33259 \, (3)$ at 15~K; see Fig.~\ref{Neutron_results}(a) for the diffraction pattern. For comparison, the corresponding room temperature values obtained in our D2B measurements are $a_{\rm lat}=10.568\,(3)$~\AA\ and $x=0.33250 \, (8)$, in agreement with the literature.\cite{Kennedy97} Our Rietveld refinement revealed a slight stuffing and oxygen deficiency: with the notation Nd$_{2+y}$Sn$_{2-y}$O$_{7+\delta}$ we got $y=0.013\,(7)$ and $\delta=-0.006\,(3)$.\cite{Bertin15a}

Magnetic neutron diffraction measurements were performed with the D1B diffractometer at ILL. In Fig.~\ref{Neutron_results}(b) a {\em magnetic} diffraction diagram at 60~mK is presented. The Bragg reflections imply a long-range order of the Nd$^{3+}$ magnetic moments to be established. The reflections only occurring at the nuclear Bragg peak positions, the magnetic propagation vector of the structure is ${\bf k} = (0,0,0)$. The absence of magnetic intensity at positions (111), (200), and (400) is consistent with the non-coplanar all-in--all-out moment arrangement pictured in Fig.~\ref{Neutron_results}(c).\cite{Bertin15a} It is confirmed by an analysis with FullProf. The spontaneous magnetic moment is $m_{\rm sp}(T \rightarrow 0) \simeq m_{\rm sp}(T = 0.06 \, {\rm K}) = 1.708 \, (3) \, \mu_{\rm B}$. The all-in--all-out magnetic structure was first reported for FeF$_3$,\cite{Ferey86} and subsequently observed for the Nd$^{3+}$ moments and proposed for the Ir moments of Nd$_2$Ir$_2$O$_7$ (Ref.~\onlinecite{Tomiyasu12a}) and for the Os moments of Cd$_2$Os$_2$O$_7$.\cite{Yamaura12,Tardif15} It was also deduced from an analysis of resonant x-ray diffraction data obtained for Eu$_2$Ir$_2$O$_7$.\cite{Sagayama13} This structure should not give rise to a structural distortion,\cite{Sagayama13} consistent with the second order nature of the magnetic phase transition. We finally mention that no diffuse scattering which would signal the onset of short-range magnetic correlations is observed when approaching the magnetic transition from the paramagnetic phase, unlike in, e.g., Tb$_2$Sn$_2$O$_7$ (Ref.~\onlinecite{Mirebeau05}); see Fig.~\ref{Neutron_results}(d).

\section{Inelastic neutron scattering results}
For an independent estimate of $m_{\rm sp}$ and to gather information on the spin dynamics up to the $10^{-9}$~s range, we performed neutron backscattering measurements at the IN16 backscattering spectrometer at ILL. 

\begin{figure}
\begin{picture}(255,98)
\put(103,78){(a)}
\put(0,0){
\includegraphics[scale=0.42]{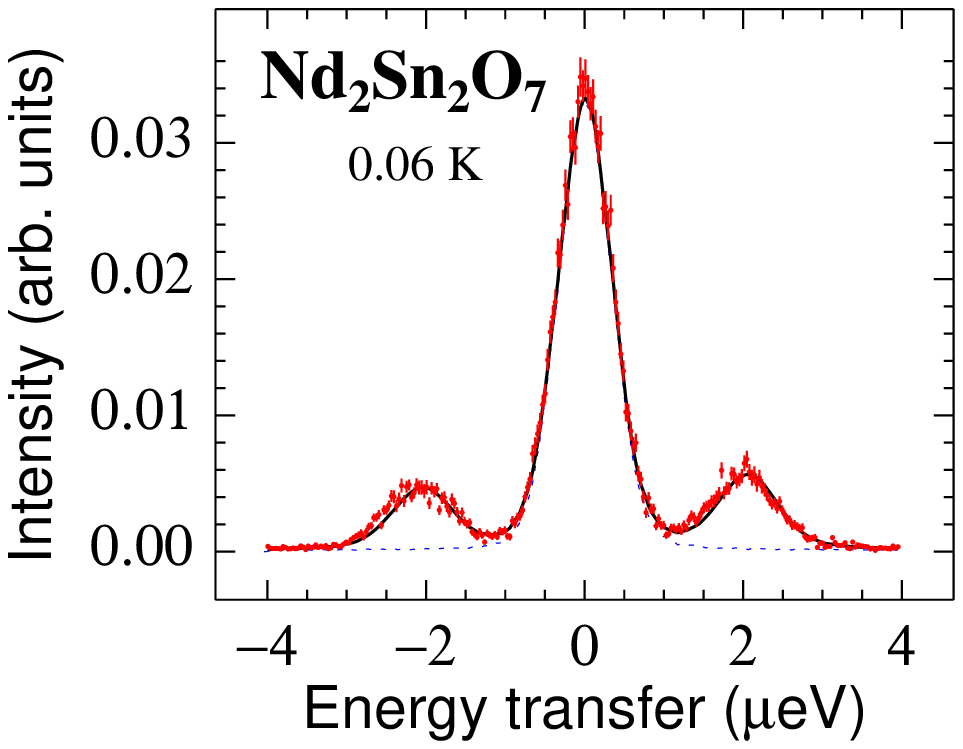}
}
\put(230,78){(b)}
\put(127,0){
\includegraphics[scale=0.42]{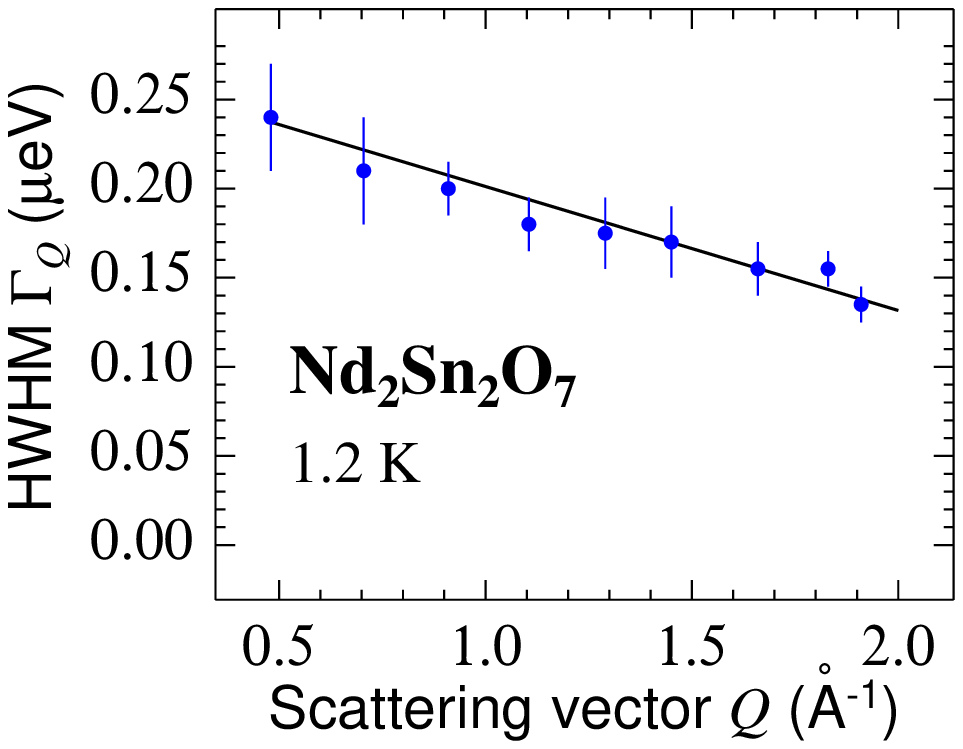}
}
\end{picture}
\caption{(Color online) Inelastic neutron scattering results for a Nd$_2$Sn$_2$O$_7$ powder sample.
(a) Wavevector integrated backscattering spectrum at 0.06~K. The instrument resolution determined from an independent measurement on vanadium is given by the dotted line.
(b) Wavevector dependence of the quasielastic half-width at half-maximum (HWHM) $\Gamma_Q$ of the magnetic scattering in the paramagnetic phase at 1.2~K. 
In the two panels, the solid lines are results of fits as explained in the main text.
} 
\label{Neutron_results_2}
\end{figure}

The inelastic scattering intensity measured at $\approx \pm 2$~$\mu$eV [Fig.~\ref{Neutron_results_2}(a)] in the magnetically ordered phase corresponds to transitions among the nuclear spin levels whose degeneracy is lifted by the hyperfine field.\cite{Heidemann70} This is an incoherent scattering process. Among the Nd isotopes the contribution of $^{143}$Nd overwhelmingly dominates due to its large incoherent scattering cross section.
The splitting $\hbar\omega_{\rm Z} = 2.027 \, (7)$~$\mu$eV measured at 0.06~K corresponds to 
$m_{\rm sp} = \hbar\omega_{\rm Z}/{\mathcal A}_{\rm hyp}^{143} = 1.68 \, (3)~ \mu_{\rm B}$ using the hyperfine constant ${\mathcal A}_{\rm hyp}^{143} = 20.9 \, (3)$~mT.\cite{Barberis79}
The incoherent nature of the scattering process implies that each nucleus is probed additively, i.e., there are no interference effects as in diffraction. Hence,  backscattering provides a {\em local probe determination} of  $m_{\rm sp}$. Diffraction, which measures a volume average,\cite{Marshall71} and backscattering data giving consistent results, no phase segregation occurs in our sample.

Information about spin dynamics is obtained from backscattering data recorded at 1.2~K, i.e., above $T_{\rm c}$. Beside the nuclear incoherent scattering, a signal of electronic origin is observed which is broader than the experimental resolution. Fitting this signal using a Lorentzian function we deduce the quasielastic half-width at half-maximum $\Gamma_Q$. We find it to be weakly $Q$-dependent with a linear wave-vector dependence $\Gamma_Q = \Gamma_0 + a_Q Q$ where $\Gamma_0= 0.271 \, (9)~\mu{\rm eV}$ and $a_Q = -0.070 \, (2)~\mu$eV\,\AA, see Fig.~\ref{Neutron_results_2}(b). While more data would be needed to discuss this $Q$ dependence, we notice the linewidth to be in the range $\Gamma_{\rm BS} \simeq$ 0.2~$\mu$eV which corresponds to a fluctuation time $\tau_{\rm BS} = \hbar/\Gamma_{\rm BS} = 3 \times 10^{-9}$\,s. This is relatively slow for a temperature outside the critical regime. We would have expected a time of order $\hbar/ (k_{\rm B} |\theta_{\rm CW}|) = 2.4 \, (1)\times 10^{-11}$~s, where we take $\theta_{\rm CW}$ derived from the $\chi (T)$ fit at low temperature. Even slower paramagnetic fluctuations are revealed by the $\mu$SR study discussed below. 

\section{$\mu$SR results}
\begin{figure}
\centering
\begin{picture}(255,275)
\put(188,255){(a)}
\put(30,140){
\includegraphics[width=0.35\textwidth]{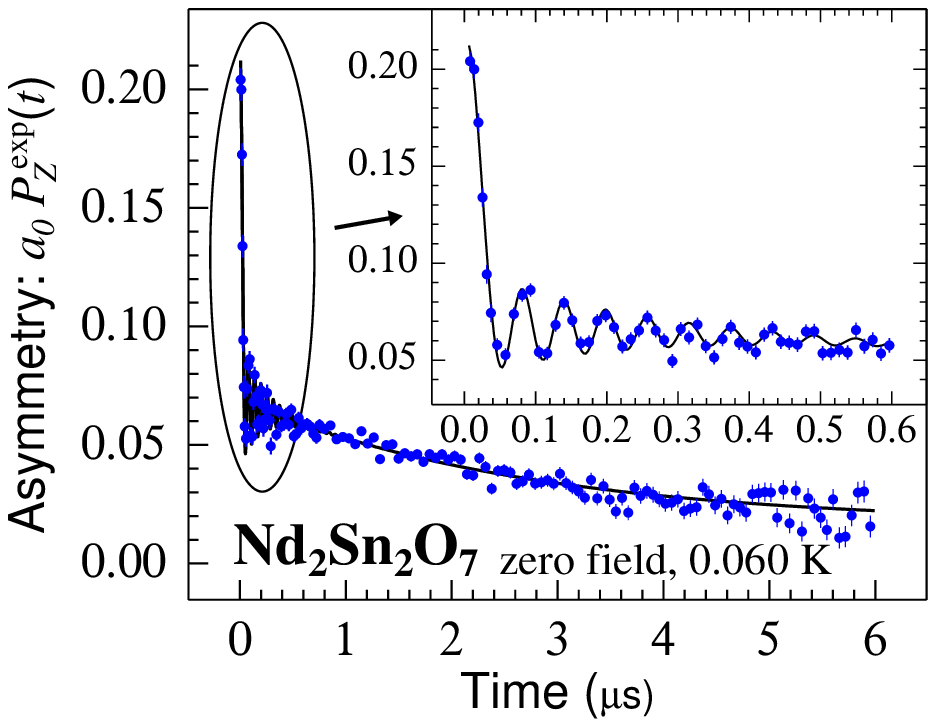}
}
\put(188,100){(b)}
\put(30,0){
\includegraphics[width=0.35\textwidth]{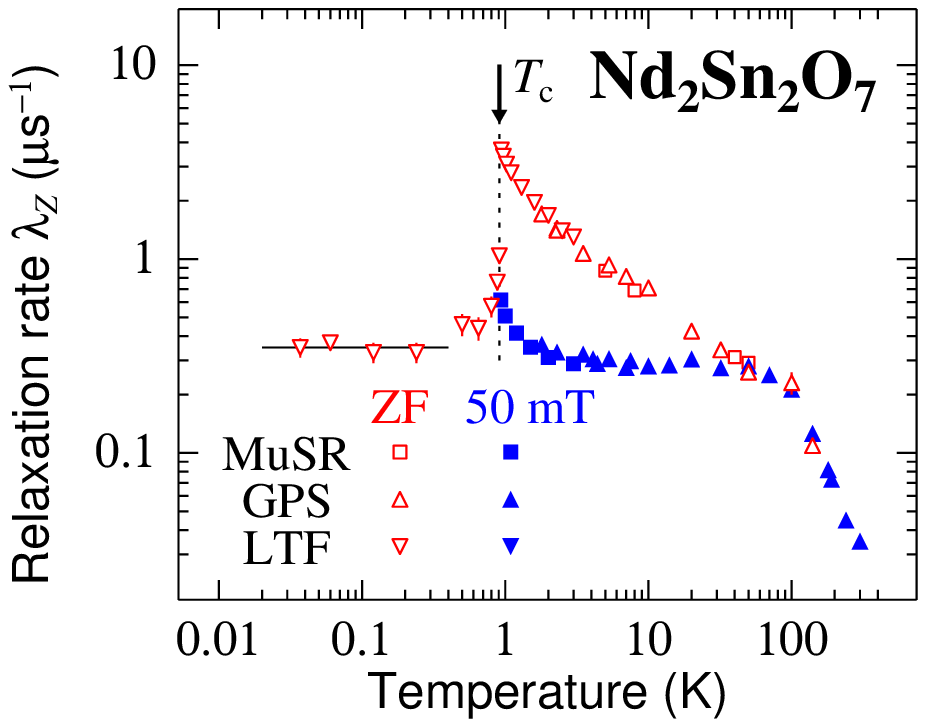}
}
\end{picture}
\caption{(Color online) $\mu$SR results.
(a) An asymmetry spectrum recorded deep in the ordered state (LTF spectrometer). The solid line results from a phenomenological fit. The insert displays the short-time part of the spectrum. (b) Temperature dependence of the spin lattice relaxation rate $\lambda_Z$ in zero field (empty symbols) and for $B_{\rm ext} = 50$~mT (full symbols). The data have been recorded at different spectrometers as indicated in the figure. The $T_{\rm c}$ value is shown as a dotted line and the full line emphasizes the temperature independent zero-field  $\lambda_Z$ at low temperature.
}
\label{speuSR}
\end{figure}

Further information was obtained from $\mu$SR measurements 
performed at the MuSR spectrometer of the ISIS pulsed muon source (Rutherford Appleton Laboratory, United Kingdom) 
and at the GPS and LTF spectrometers of the Swiss Muon Source at PSI. All the data were recorded with the longitudinal geometry, either in zero field (ZF) or under a longitudinal field. These different variants of the technique probe the magnetic properties of the system on a time scale ranging from approximately $10^{-5}$ to $10^{-12}$~s.
A $\mu$SR asymmetry spectrum recorded below $T_{\rm c}$ is shown in Fig.~\ref{speuSR}(a). 
The displayed quantity is $a_0 P^{\rm exp}_Z(t) $ where $a_0$ is an experimental parameter and $P^{\rm exp}_Z(t)$ is the muon polarization function which reflects the physics of the compound under study.\cite{Yaouanc11} We observe a sharp drop of the asymmetry at short times followed by a slow exponential decay of the tail. Its decay rate is a measure of the spin-lattice relaxation rate $\lambda_Z$. The early time part of the spectrum is detailed in the insert. A spontaneous, i.e., in the absence of an external field, muon spin precession is detected. This is so up to 
$T \leq 0.65~{\rm K} \approx 0.7 ~T_{\rm c}$. 
It reflects the long-range magnetic order. Although a spontaneous muon spin 
precession is expected and often observed in the ordered phase of magnets, including exotic magnets,\cite{Yaouanc08} it is not present, as mentioned earlier, for Tb$_2$Sn$_2$O$_7$,\cite{Dalmas06,Bert06} Er$_2$Ti$_2$O$_7$,\cite{Lago05,Dalmas12a} and Yb$_2$Sn$_2$O$_7$,\cite{Yaouanc13,Lago14} and for CdHo$_2$S$_4$.\cite{Yaouanc15} This has been attributed to the dynamical nature of the field at the muon site.\cite{Dalmas06} Above $\approx 0.7 ~T_{\rm c}$, the muon spin precession is not resolved: this is probably 
due to a broadening of the field distribution at the muon site and dynamical effects. In the whole temperature 
range below $T_{\rm c}$ the spin-lattice relaxation channel is easily observed. 
The function $a_0 P^{\rm exp}_Z(t) = a_{\rm s} P_Z(t) + a_{\rm bg}$ was fitted to all 
the spectra. The first term is associated with muons probing the sample and the second time-independent 
component accounts for muons implanted in the sample surroundings. 
Above $T_{\rm c}$ a stretched exponential function $P_Z(t) = \exp[-(\lambda_Z t)^{\beta_{\rm se}}]$, 
with the exponent $ 0.7 \leq \beta_{\rm se} \leq 1$, provides a good description of the spectra. Such a spectral shape is consistent with a distribution of relaxation channels and is often observed in frustrated magnets.

In Fig.~\ref{speuSR}(b), $\lambda_Z(T)$ deduced from zero-field and 50~mT measurements is plotted. 
These data are remarkable in many respects. 
In zero field $\lambda_Z(T)$
displays a pronounced maximum at $T_{\rm c}$. This reflects the slowing down of the critical fluctuations at the approach of a second order magnetic phase transition. At $T \ll T_{\rm c}$ 
we would expect $\lambda_Z$ to vanish.\cite{Yaouanc91,Dalmas95,Dalmas04,Yaouanc11}
The plateau that is instead observed, the signature of the so-called persistent spin dynamics, has been ascribed to unidimensional  
excitations\cite{Yaouanc15} which could be present in lattices of corner-sharing triangles or tetrahedra. The $\lambda_Z$ thermal behavior above its inflection point 
at $\approx$~100~K is driven by an Orbach local relaxation 
mechanism.\cite{Dalmas03,Yaouanc11} At lower temperature, but still in the
paramagnetic regime, correlations come into play. However, their characteristics is not as expected for a conventional paramagnetic phase. This key new result is dramatically  revealed by the strong dependence
of $\lambda_Z$ on $B_{\rm ext}$.  
Quite unexpectedly, its influence extends up to about 30~K, i.e., $\approx 30 \, T_{\rm c}$. Because of this strong $B_{\rm ext}$ dependence of $\lambda_Z$ at low field and the plateau observed for $\lambda_Z(T)$ under 50~mT for $2 < T < 100$~K, using the model due to Redfield\cite{Redfield57} we infer the presence of spin fluctuations  with a correlation time $\tau_{\mu{\rm SR}}$ in the 100~ns range. Hence, the zero-field fluctuations probed by $\mu$SR are characterized by $\tau_{\mu {\rm SR}}$ much larger than the time estimated from the quasielastic neutron scattering data, i.e., $\tau_{\rm BS}$. 
Nd$_2$Sn$_2$O$_7$ is not a unique example of this feature.\cite{Gardner04}
In fact, a wide range of correlation times extending to anomalously long times seems to be a signature of geometrically 
frustrated magnetic materials.\cite{Ueland06,Chapuis07,Rule09b,Dalmas12a,Dalmas12,Legros15,Maisuradze15,Nambu15}

\section{Discussion and Conclusions}
We now discuss the prominent experimental results obtained from our experimental study, namely (i) the all-in--all-out type of magnetic structure found for Nd$_2$Sn$_2$O$_7$, (ii) the observation of a spontaneous field in $\mu$SR measurements while this field is not detected in some other magnetically ordered pyrochlores, and (iii) the anomalously slow dynamics in the paramagnetic phase.

The all-in--all-out magnetic structure has been predicted for 
antiferromagnetically coupled Ising spins.\cite{Bramwell98} However, the $T^3$ temperature dependence 
of the specific heat implies that magnon-like excitations are present and therefore the spins cannot
be entirely Ising-like. A recent computation using a gauge mean-field approximation and incorporating 
spin exchange terms other than the Ising term has appeared.\cite{Huang14}
It predicts the all-in--all-out phase for a range of exchange parameters. Unfortunately this result 
cannot be tested since no formulas for the velocity $v_{\rm ex}$ and the static susceptibility 
$\chi$ in terms of exchange parameters have been given. 

Recently the concept of fragmentation for Ising anisotropy pyrochlore compounds has been introduced.\cite{Brooks14}
The field associated with the magnetic moments of an isolated tetrahedron is proposed to split into divergence-full
and divergence-free parts. Within this model a long-range order is modelled by the first part
of the Helmholtz decomposition. The second part corresponds to a fluid of monopoles. Since the divergence-free
field does not exist for the ground state of the all-in--all-out magnetic structure, there is no reason for the spontaneous
field to flip\cite{Dalmas06} and therefore its detection for Nd$_2$Sn$_2$O$_7$ is naturally explained.   
We note that the low temperature moment 
is not collinear with the local three-fold axis
for Tb$_2$Sn$_2$O$_7$, Yb$_2$Sn$_2$O$_7$, and Yb$_2$Ti$_2$O$_7$. Hence the fragmentation model in its present status is not applicable to these three compounds. Notice that the physics of Tb$_2$Sn$_2$O$_7$ is in any case complicated because Tb$^{3+}$ is a non-Kramers doublet and therefore the lattice is subject to crystallographic distortions.

Slow paramagnetic dynamics might be a general property of magnetic pyrochlore compounds. It was uncovered early on for Tb$_2$Ti$_2$O$_7$.\cite{Ueland06} It has been recently characterized in Yb$_2$Ti$_2$O$_7$ and Yb$_2$Sn$_2$O$_7$.\cite{Maisuradze15} As to its origin, emergent unidimensional spin clusters --- spin loops --- i.e., peculiar molecular spin structures, are a natural candidate. Excitations supported by these structures were recently suggested to drive persistent spin dynamics for $T \ll T_{\rm c}$ as measured by $\lambda_Z$.\cite{Yaouanc15} The interactions between the quasiparticles above $T_{\rm c}$ may renormalize them which would lead to the observed temperature dependence of $\lambda_Z$. Spins involved in these structures are expected to display short-range correlation effects and their dynamics should be relatively slow in comparison to magnetically isolated spins. These structures have rarely been discussed theoretically.\cite{Villain79,Hermele04,Yavors08} They are of natural occurrence in pyrochlore and kagome lattices. 

While spin dynamics has been consistently observed in numerous studies of the pyrochlore systems, the present work suggests that it signals phenomena of quite different origins. The temperature independent muon-spin relaxation rate when $T\to 0$ and the anomalously slow dynamics revealed here are attributed to correlated spin loops. The absence of a spontaneous field would result from the divergence-free part of the magnetization field in compounds with components of their magnetic moments along local three-fold axes.

For further progress, the published fragmentation model, which only considers classical Ising 
spins,\cite{Brooks14} should be generalized to quantum spins with components perpendicular to the local 
three-fold axis.\cite{Curnoe08} The generic nature of the unidimensional excitations should be addressed. 
On the experimental front, extending the study of insulating corner-sharing regular tetrahedra systems to normal rare-earth spinel 
systems is poised to provide more examples of exotic spin dynamics as recently shown.\cite{Lago10,Yaouanc15} 

\begin{acknowledgments}
We are grateful to  P.C.W. Holdsworth and E. Ressouche for discussions.
AY acknowledges speakers at the International Workshop on Frustration
and Topology in Condensed Matter Physics held at Tainan in February 2014 for clarifying
the nature of excitations in geometrically frustrated compounds. He thanks
S. Onoda for his invitation to the workshop.
This research project was partially supported by the European Commission 
under the 6th Framework Programme through the Key Action: Strengthening the European Research 
Area, Research Infrastructures (Contract number: RII3-CT-2003-505925) and 
under the 7th Framework Programme through the `Research Infrastructures' action of the
`Capacities' Programme, Contract No: CP-CSA\_INFRA-2008-1.1.1 Number
226507-NMI3. Part of this work was performed at the Institut Laue Langevin, Grenoble, France, at the muon and neutron sources of the Paul Scherrer Institute, Villigen, Switzerland and at the ISIS pulsed muon facility, Rutherford Appleton Laboratory, Chilton, United Kingdom.
\end{acknowledgments}

\bibliography{reference,Nd2Sn2O7_prb_footnote.bib}

\end{document}